\documentstyle[12pt]{article}

\newcommand{\beq}{\begin{equation}}
\newcommand{\eeq}{\end{equation}}
\newcommand{\beqa}{\begin{eqnarray}}
\newcommand{\eeqa}{\end{eqnarray}}

\begin{document}
\topmargin 0pt
\oddsidemargin 1mm
\begin{titlepage}
\begin{flushright}
DSFTH 11/98\\
June, 1998\\
\end{flushright}
\setcounter{page}{0}
\vspace{15mm}
\begin{center}
{\Large   ON OFF-SHELL BOSONIC STRING AMPLITUDES } 
\vspace{20mm}

{\large Luigi Cappiello}, {\large Raffaele Marotta}\\
{\large Roberto Pettorino} and {\large Franco Pezzella~\footnote{email:
Name.Surname@na.infn.it}}\\

{\em Dipartimento di Scienze Fisiche, Universit\`{a} di Napoli\\
and I.N.F.N., sezione di Napoli\\
Mostra d'Oltremare, pad. 19, I-80125 Napoli, Italy}\\
\end{center}
\vspace{7mm}

\begin{abstract}
We give a simple prescription for computing, in the framework of the bosonic
string theory, off-shell one-loop amplitudes with any number of external
massless particles, for both the open and the closed string. We discuss their
properties and, in particular, for the two-string one-loop amplitudes we show
their being transverse.
\end{abstract}

\vspace{1cm}

\end{titlepage}
\newpage
\renewcommand{\thefootnote}{\arabic{footnote}}

Perturbative string theories rely on prescriptions for computing
scattering amplitudes involving on-shell physical external states. 

String amplitudes are expressed as an integral over the moduli 
space of a punctured Riemann surface and 
the integrand is a correlator of vertex operators corresponding to
the external states inserted at the punctures. 

In on-shell string amplitudes the vertex operators are always primary fields
of the underlying conformal field theory and it can be 
shown~\cite{DPFHLS} \cite{NW} that such amplitudes are independent of the
choice of a local coordinate system at the punctures. This choice is
instead crucial for off-shell string amplitudes. In this case
the vertex operators are non-primary fields and their
correlators turn out to be dependent on which local holomorphic
coordinate system is used. 
  
Off-shell continuations have been studied a great deal until now 
\cite{CMNP}$\div$\cite{DMLRM}
and different prescriptions have been given according to
the pursued approaches.

There are important reasons for studying off-shell amplitudes,
firstly the low energy limit 
($\alpha' \rightarrow 0$) of string theories. This limit    
has to reproduce perturbative aspects of
the ordinary gauge field theories describing the fundamental interactions,
including gravity. Moreover it provides a very powerful computational
tool: instead of calculating field theory amplitudes
by the conventional Feynman diagrams, it is easier to calculate
the corresponding string ones, that have a more compact expression and
are much fewer, and then perform their low-energy limit \cite{BK}
\cite{DMLRM}. In particular
closed strings can be used to shed light on perturbative quantum
gravity and its divergencies \cite{BDS}. 
In fact, perturbative computations in gravity are known to be
algebraically very complex and calculations of gravity
amplitudes with more than two loops \cite{GS}
\cite{V} have never been performed. 

In this respect it  could be very interesting to analyse the $ \alpha'
\rightarrow 0$ limit of closed string amplitudes for massless states
in the one-loop case. Indeed for the two-point function one
should be able to make a comparison with the well-known one-loop 
counterterms in quantum gravity \cite{HV}. Moreover we can use 
the purely bosonic string since we can deal with the tachyonic divergences 
as in the ref. \cite{DMLRM}.

Two steps are needed: first, to give a reliable prescription for
off-shell closed string amplitudes; second, to consider, at one loop
level, the sum of the amplitude on the torus and the one on the Klein
bottle, in order to reproduce results in a theory where the antisymmetric
tensor is decoupled from the graviton and the dilaton.

These are our main motivations; however it is by now evident that there
are also reasons of interest in off-shell string amplitudes {\em per
se} as, for instance, processes 
involving closed string exchange among D-branes suggest
\cite{P}  \cite{RT} \cite{BS}. 

In this letter, we limit ourselves to the first step indicated above, 
and, in the framework of the operatorial formalism,
we give a general prescription for computing off-shell 
one-loop string amplitudes with any number $N$ of external massless particles.
We find  a simple choice of the local coordinates at the punctures  
on the Riemann surface, such that {\em none} of the on-shell conditions
on the external states has to be kept. The string
amplitude obtained in the case $N=2$ turns out to satisfy the 
transversality condition.  

Our starting point is the $N$-string $g$-loop Vertex for the oriented bosonic
closed string \cite{DPFHLS}:

\[
V_{N;g}= {\cal C}_{g}\prod_{i=1}^{N} \bigg[ {}_i\!<\!
p_{i};O_{a},O_{{\bar{a}}}|\bigg] \,\,\delta  \left( \sum_{i=1}^{N} p_{i} 
\right)
\int [{\mbox d}m]^{N}_{g}
\,{\hat{V}}_{N;0}
{\hat{\bar{V}}}_{N;0}
\]
\[
\times \exp\left\{-\frac{1}{2}\sum_{i,j=1}^{N}
\oint_0 dz \oint_0 dy \partial X^{(i)}(z)
\log
\frac{ E\left(V_i(z),V_j(y)\right)}
{V_{i}(z)-V_{j}(y)}
\partial X^{(j)}(y)
\right\}
\]
\[
\times \exp\left\{-\frac{1}{2}\sum_{i,j=1}^{N}
\oint_0 d{\bar{z}} \oint_0 d{\bar{y}} {\bar{\partial}}
{\bar{X}}^{(i)}(\bar z)
\log
\frac{ E\left({\bar{V}}_i(z),{\bar{V}}_j(y)\right)}
{{\bar{V}}_{i}(z)-{\bar{V}}_{j}(y)}
{\bar{\partial}} {\bar{X}}^{(j)}(\bar y)
\right\}
\]
\[
\times \int \prod_{\mu=1}^{g} [ d^{D} k_{\mu} ] \exp\left\{
-\alpha' \sum_{\mu,\nu=1}^{g} k_{\mu} ( \pi {\mbox Im} \tau )_{\mu\nu} k_{\nu}
\right\}
\]
\beq
\times \exp\left\{i \sqrt{\frac{\alpha'}{2}} \sum_{i=1}^{N} 
\sum_{\mu=1}^{g}k_{\mu}
\left[ \oint_0 dz \partial X^{(i)}(z)
\left(
\int_{z_{0}}^{V_{i}(z)} \omega^{\mu} \right) +
\oint_0 d{\bar{z}} {\bar{\partial}} {\bar{X}}^{(i)}(\bar z)
\left(
\int_{{\bar{z}_{0}}}^{{\bar{V}}_{i}(z)} {\bar{\omega}}^{\mu} \right) \right]
\right\}
\label{VNG}
\eeq
where ${\hat{V}}_{N;0}$ is given by
\[
{\hat{V}}_{N;0}=
\exp\left\{-\frac{1}{2}
{\sum_{\stackrel{i,j=1}{i\neq j}}^{N}}\oint dz
\oint dy \partial X^{(i)}(z) \log[V_i(z) - V_j(y)] \partial
X^{(j)}(y)\right\}
\]
\beq
\times \exp \left\{\frac{i}{2} \oint dz \partial X^{(i)}(z) \alpha_0^{(i)} \log
V_i'(z) \right\}
\label{VN0}
\eeq

with an analogous definition for $\hat{\bar{V}}_{N;0}$. 
In (\ref{VNG}) $\left[ {\mbox d}m \right]^{N}_{g}$ is the measure of
integration on the moduli space for a closed Riemann surface of genus $g$.
It includes the whole ghost contribution~\cite{DFLS}. ${\cal C}_{g}$ is a
normalization constant depending on the genus of the surface \cite{CMR}.                   
The fundamental lenght $\sqrt{2 \alpha'}$ has been extracted from
the string coordinates in order to introduce the dimensionless
fields:

\beq
X^\mu (z,\bar{z})=\frac{1}{2} \left[ X^\mu (z)+\bar{X}^\mu ( 
\bar{z}) \right]                    
                                             \label{XZZ} 
\eeq

with

\beqa
X^{\mu} (z) = q^\mu -i \alpha_{0}^{\mu}  \log z\
+i\sum\limits_{n\neq 0}\frac{\alpha_n^\mu}{n}z^{-n} \\
\bar{X}^{\mu} (\bar{z}) = q^{\mu} - i \bar{\alpha}_{0}^{\mu}
\log \bar{z} + i\sum\limits_{n\neq 0}\frac{\bar{\alpha}^{\mu}_{n}}{n}
\bar{z}^{-n} 
                            \label{XSZ}
\eeqa           
being $\alpha_{0}$ and $\bar{\alpha}_{0}$ defined by:

\[
\alpha_{0}^{\mu} = \bar{\alpha}_{0}^{\mu} =
\sqrt{\frac{\alpha^{\prime}}2} p^{\mu}  .
\]

In the operator formalism $V_{N;g}$, as defined in Eq.~(\ref{VNG}), provides
an efficient way of computing scattering amplitudes among arbitrary states
at all orders in perturbation theory. In fact, by saturating the operator
$V_{N;g}$ with $N$ external states $ |\alpha_{1}> \cdots | \alpha_{N}> $,
the corresponding amplitude is obtained:
\begin{equation}
A^{(g)}(\alpha _1,\ldots \alpha_N)=V_{N;g}|\alpha_1> \cdots
|\alpha_N> .     \label{GAM} 
\end{equation}

The Vertex $V_{N;g}$ depends on the $N$ complex Koba-Nielsen variables
$z_{i}$'s corresponding to the punctures of the external states, 
through $N$ 
conformal transformations $V_{i}(z)$'s , which define a local coordinate 
system vanishing around each $z_{i}$, i.e.:
\beq
               V_{i} (0) = z_{i}. \label{G1}
\eeq

When $V_{N;g}$ is saturated with $N$ physical string states satisfying the
mass-shell condition, the corresponding amplitude does not depend on the
$V_{i}$'s. If this condition is relaxed, the dependence of
$V_{N;g}$ on them is transferred
to the off-shell amplitude. This is analogous to what happens
in gauge theories, where on-shell amplitudes are gauge invariant, while
their off-shell counterparts are not.

Let us concentrate on the case in which all the $N$ external states belong
to the massless level of the bosonic closed string; let 
$|\epsilon; p>$ be such a state with polarization $\epsilon$ and momentum $p$:

\begin{equation}
|\epsilon; p>={\cal N}_0\epsilon _{\mu \nu}\alpha _{-1}^\mu \overline{\alpha }%
_{-1}^\nu |p> . \label{MS0}
\end{equation}
Depending on the symmetry properties of the polarization $\epsilon
_{\mu \nu}$, this state defines the antisymmetric tensor or a combination of
graviton and dilaton states. In our notations the normalization factor  
is ${\cal N}_0=\kappa /\pi $, where $\kappa$ is the gravitational 
constant in $d$ dimensions.

An on-shell massless state satisfies the conditions:
\beqa
p^2=0 &{~~~~~~~~~~}&\epsilon \cdot
  p =0  \label{onshell}
\eeqa
and it is defined through its corresponding vertex operator, which
is a primary field of the underlying conformal field theory:
\begin{equation}
|\epsilon; p>=\lim _{z,\bar{z}\rightarrow 0}V(z, 
\bar{z})|0>    \label{O-S}
\end{equation}

with
\begin{equation}
V(z,\bar{z})=i {\cal N}_0 \epsilon _{\mu \nu}:\partial_z X^\mu
(z)\partial _{\bar{z}}\bar{X}^\nu (\bar{z})e^{i\sqrt{2\alpha
^{\prime }}p\cdot X(z,\bar{z})}:         \label{VER}
\end{equation}

Since we are interested in off-shell amplitudes, we will release 
the conditions (\ref{onshell}) and  
we will write down the
expression of $V_{N;g}$ ready to be saturated with such states; we
denote it as ${\cal V}_{N;g}$ and it is given by:
\[
{\cal V}_{N;g}\, =\, {\cal C}_{g} < \Omega| \int [ {\mbox d} m]^{g}_{N}
\exp \left\{\frac{1}{2} \sum_{i=1}^{N} \sqrt{\frac{\alpha'}{2}} p^{(i)} \cdot
\left.\! \left[ \sqrt{\frac{\alpha'}{2}} p^{(i)} + {\alpha}_{1}^{(i)} 
\partial_{z}
+ \bar{\alpha}_{1}^{(i)} \partial_{\bar{z}} \right] \log |V'_{i}(z)|^2
\right|_{z=0} \right\}
\]
\[
\times \exp  \bigg\{  \sum_{\stackrel{i,j=1}{i \neq j}}^{N} 
\left[ \sqrt{\frac{\alpha'}{2}} p^{(i)}  + \alpha_{1}^{(i)} V'_{i}(0)
\partial_{z_{i}}  + \bar{\alpha}_{1}^{(i)} \bar{V}'_{i}(0) 
\partial_{\bar{z}_{i}} 
\right] 
\]
\[\cdot
\left. 
\left[ \sqrt{\frac{\alpha'}{2}} p^{(j)}  + \alpha_{1}^{(j)}
V'_{j}(0) \partial_{z_{j}}  + \bar{\alpha}_{1}^{(j)} \bar{V'}_{j}(0)
\partial_{\bar{z}_{j}}  \right]  {\cal G}(z_{i},z_{j} ) \right\}
\]
\beq
\times \exp \left\{  -2 \sum_{i=1}^{N} \alpha_{1}^{(i)} \cdot 
\bar{\alpha}_{1}^{(i)} |V'_{i}(0)|^{2} \partial_{z} \partial_{\bar{z}} 
\,\,{\cal G}(z,z_{i} )\left.  \right|_{{z=z_{i}}} \right\}
\label{VML}
\end{equation}
where
\beq
{\cal G}(z_{i},z_{j}) \equiv \frac{1}{2} {\log} |E(z_{i},z_{j})|^{2}
- \frac{1}{2}  \left(
{\mbox Re} \int_{z_{i}}^{z_{j}} \omega^{\mu} \right) ( 2 \pi {\mbox Im} \tau
 )_{\mu \nu}^{-1} \left( {\mbox Re} \int_{z_{i}}^{z_{j}} \omega^{\nu} 
\right)           \label{GF}
\eeq
is the $g$-loop world-sheet bosonic Green function \cite{GSW}.

The expression of ${\cal V}_{N;g}$ in (\ref{VML}) has been 
obtained by performing 
the gaussian integration over the
internal momenta $k_{\mu}$'s.
It has generated the term, written in the last line,
in which right and left movers are mixed; in other words, the vertex 
is $not$ holomorphically factorized.

We can exploit the freedom of choosing the
conformal local maps $V_{i}(z)$'s in order to write ${\cal V}_{N;g}$ 
in a more geometrical form depending only on the Green function. First we
rescale this latter as follows:
\beqa
G(z_{i}, z_{j})& = & {\cal G}(z_{i},z_{j}) - \frac{1}{4} \log |V'_{i}(0)
V'_{j}(0)|^{2} \nonumber  \\
 & = & \frac{1}{2} \log \frac{|E(z_{i},z_{j})|^{2}}{|V'_{i}(0)V'_{j}(0)|}
- \frac{1}{2}  \left(
{\mbox Re} \int_{z_{i}}^{z_{j}} \omega^{\mu} \right) ( 2 \pi {\mbox Im} \tau
 )_{\mu \nu}^{-1} \left( {\mbox Re} \int_{z_{i}}^{z_{j}} \omega^{\nu} 
\right)     
\eeqa
This expression coincides with the one given in literature~\cite{M}
\cite{RS}, where the conformal maps $V_{i}(z)$'s may depend on
all the moduli of the world sheet and on the positions of 
the punctures.  

In the following we restrict ourselves to the $g=1$ loop
case and, at  this order, a choice corresponding to
\beq
V'_{i}(0)= z_{i} \label{G2}   
\eeq
reproduces the translational invariant Green function on the 
torus \cite{GSW}.
However a careful inspection of the expression (\ref{VML}) 
shows that the condition (\ref{G2}) is not sufficient to write the vertex 
${\cal V}_{N;1}$ only in terms of the Green function; we are then led 
to impose the following further condition:
\beq
V''_{i}(0)=z_{i}  \label{G3}
\eeq
The constraints (\ref{G1}), (\ref{G2}) and (\ref{G3}) may be satisfied,
choosing, for instance, the following holomorphic local coordinate map
at the puncture $z_{i}$ :
\beq
V_{i}(z) = z_{i} e^{z} .  \label{GAUGE}
\eeq
This choice allows us to rewrite ${\cal V}_{N;1}$ as:
 
\[
{\cal V}_{N;1}\, =\, {\cal C}_{1}\! < \Omega| \int [ {\mbox d} m]^{1}_{N}
\exp \bigg\{ \sum_{\stackrel{i,j=1}{i \neq j}}^{N} 
\left[ \sqrt{\frac{\alpha'}{2}} p^{(i)} + \alpha_{1}^{(i)} z_{i}
\partial_{z_{i}} + \bar{\alpha}_{1}^{(i)} \bar{z}_{i} \partial_{\bar{z}_{i}}
\right] 
\]
\[
\cdot  \left[ \sqrt{\frac{\alpha'}{2}} p^{(j)} + \alpha_{1}^{(j)}
z_{j} \partial_{z_{j}} + \bar{\alpha}_{1}^{(j)} \bar{z}_{j}
\partial_{\bar{z}_{j}} \right]  G(z_{i},z_{j} ) \bigg\}
\]
\beq
\times \exp \left\{  -2 \sum_{i=1}^{N} \alpha_{1}^{(i)} \cdot 
\bar{\alpha}_{1}^{(i)} |z_{i}|^{2} \partial_{z} \partial_{\bar{z}} 
\,\, G(z,z_{i} )\left.  \right|_{{z=z_{i}}} \right\}
\label{VML1}
\end{equation}
and it reproduces, for small values of $z$,
the gauge 
\beq
V_{i}(z) = z_{i}z + z_{i}      \label{GAUGEOP}
\eeq
that has been proposed for the open string~\cite{DMLRM}. 
We point out, however, that in this case, the map
(\ref{GAUGEOP}) has to be considered together with the condition
$\epsilon \cdot p = 0$ for external photons even if off mass-shell. 
Differently, our choice (\ref{GAUGE}) 
does not need to be coupled to any other  condition:  
our proposal for the local maps $V_{i}(z)$'s includes {\em all} the off-shell
conditions. In the following we will check for the two-gluon amplitude
that this prescription also works in the open string case.

The vertex ${\cal V}_{N;1}$ (\ref{VML1}) is a very closed expression that 
gives the possibility of easily computing off-shell one-loop amplitudes for an 
arbitrary number of external massless states. Due to the structure of the
closed string states (\ref{MS0}), one can think at the polarization
tensor as: 
\[
 \epsilon_{\mu \nu} = \xi_{\mu} \otimes \bar{\xi}_{\nu} .  
\]
Therefore it is straightforward writing the $N$-string one-loop
amplitude:
\[
A_{N;1}\, =\, {\cal C}_{1}\! \int [ {\mbox d} m]^{1}_{N}
\exp \bigg\{ \sum_{\stackrel{i,j=1}{i \neq j}}^{N} 
\left[ \sqrt{\frac{\alpha'}{2}} p^{(i)} + \xi^{(i)} z_{i}
\partial_{z_{i}} + \bar{\xi}^{(i)} \bar{z}_{i} \partial_{\bar{z}_{i}}
\right] 
\]
\[
\cdot  \left[ \sqrt{\frac{\alpha'}{2}} p^{(j)} + \xi^{(j)}
z_{j} \partial_{z_{j}} + \bar{\xi}^{(j)} \bar{z}_{j}
\partial_{\bar{z}_{j}} \right]  G(z_{i},z_{j} ) \bigg\}
\]
\beq
\times \exp \left\{  -2 \sum_{i=1}^{N} \xi^{(i)} \cdot 
\bar{\xi}^{(i)} |z_{i}|^{2} \partial_{z} \partial_{\bar{z}} 
\,\, G(z,z_{i} )\left.  \right|_{{z=z_{i}}} \right\} .
\label{AML1}
\end{equation}
This expression has to be understood as an expansion in $\xi^{(i)}$
and $\bar{\xi}^{(i)}$ restricted only to the linear terms. 

In order to check the properties of the off-shell amplitudes obtained
through the vertex~(\ref{VML1}), we consider the case $N=2$.

The normalization factor is given by ${\cal C}_{1}= \left(2 \pi
\alpha^{\prime} \right)^{- d/2} $ \cite{CMR}. 
The one-loop measure is

$$
\left[ {\mbox d}m \right]_2^{1}=\frac{d^{2} z_1 d^{2}z_{2}}{|z_{1}|^{2}
|z_{2}|^{2}} \,\frac{d^{2}k}{|k|^4}\left[ -\ln |k| \right]^{-d/2\,}
\prod_{n=1}^{+\infty} \left(
\left| 1-k^n\right|^2 \right)^{2-d}. 
$$

and the explicit expression of the Green function~(\ref{GF}) is \cite{GSW}:

$$
{\cal G}(z_i,z_j)=\frac{1}{2} \log  \left| (z_{i} - z_{j}) 
\prod_{n=1}^{+\infty } \frac{
(z_{i}-k^{n}z_{j})(z_{j}-k^{n}z_{i})}{z_{i}z_{j}(1-k^{n})^2}\right| ^2+
\frac{1}{2} \frac{ {\log}{}^2 |z_1/z_2| }{\log |k|} .
$$

One-loop two-string amplitudes can be obtained by expanding  $A_{2;1}$
up to terms linear in $\xi^{(1)\mu} \bar{\xi}^{(1)\nu}
\xi^{(2)\rho} \bar{\xi}^{(2)\sigma}$. After some algebra we get:

\[
A_{2;1} = {\cal N}_{0}^{2} {\cal C}_{1} \epsilon^{\mu \nu (1)}\epsilon^{\rho \sigma (2)}
                          T_{\mu \nu \rho \sigma}
\]
with

$$
\begin{array}{ll}
T_{\mu \nu \rho \sigma} =  
  &  4 \bigg[ \, \eta _{\mu \nu }\eta _{\rho \sigma }\ a_1+\eta _{\mu
\sigma }\eta _{\nu \rho }\ a_2+\eta _{\mu \rho }\eta _{\nu \sigma }\ a_3
\\  
& +\alpha ^{\prime } \bigg( \, \eta _{\mu \rho }\ p^{(1)}_{\sigma }\ p^{(2)}_{\nu \ }a_4 - \eta
_{\mu \nu }\ p^{(1)}_{\rho }\ p^{(1)}_{\sigma \ }a_5+\eta _{\mu \sigma }\
p^{(1)}_{\rho
}\ p^{(2)}_{\nu \ }a_6 \\  
& + \eta _{\nu \rho }\ p^{(1)}_{\sigma }\ p^{(2)}_{\mu \ }a_7-\eta _{\rho 
\sigma }\
p^{(2)}_{\mu }\ p^{(2)}_{\nu \ }a_8 + \eta _{\nu \sigma }\ p^{(1)}_{\rho }\ 
p^{(2)}_{\mu \
}a_9 \bigg) \\    
& + (\alpha ^{\prime })^2\ p^{(1)}_{\rho }\ p^{(1)}_{\nu \ }\
p^{(2)}_{\mu }\ p^{(2)}_{\sigma \ }a_{10} \bigg] 
\end{array}
$$
where the coefficients $a_i$'s, $i=1,\ldots,10$, are given by:

$$
\begin{array}{ll}
a_{1}= \int
 \left[ dm \right]_{2}^{1} e^{{\alpha}^{\prime} p^{(1)} \cdot p^{(2)}
G(z_{1},z_{2})}|z_{1}|^{2} |z_{2}|^{2} \partial_{z_{1}} \partial_{\bar{z}_{1}}
G(z_{1},z_{2}) \partial_{z_{2}} \partial_{\bar{z}_{2}} G(z_{1},z_{2}) \\

a_2=\int \left[ dm \right]_2^{1} e^{\alpha ^{\prime} p^{(1)} \cdot p^{(2)}
G(z_{1},z_{2}) }|z_{1}|^{2} |z_{2}|^{2} \partial_{z_{1}} \partial_{\bar{z}_{2}}
G(z_{1},z_{2}) \partial_{\bar{z}_{1}} \partial_{{z}_{2}} G(z_{1},z_{2}) \\

a_3=\int \left[ dm \right]_2^{1} e^{\alpha ^{\prime} p^{(1)} \cdot p^{(2)}
G(z_{1},z_{2}) }|z_{1}|^{2} |z_{2}|^{2} \partial_{z_{1}} \partial_{{z}_{2}}
G(z_{1},z_{2}) \partial_{\bar{z}_{1}} \partial_{\bar{z}_{2}} G(z_{1},z_{2}) \\

a_4=\int \left[ dm \right]_2^{1} e^{\alpha ^{\prime} p^{(1)} \cdot p^{(2)}
G(z_{1},z_{2}) }|z_{1}|^{2} |z_{2}|^{2} \partial_{z_{1}} \partial_{{z}_{2}}
G(z_{1},z_{2}) \partial_{\bar{z}_{1}}
G(z_{1},z_{2}) \partial_{\bar{z}_{2}} G(z_{1},z_{2}) \\

a_5=\int \left[ dm \right]_2^{1} e^{\alpha ^{\prime} p^{(1)} \cdot p^{(2)}
G(z_{1},z_{2}) }|z_{1}|^{2} |z_{2}|^{2} \partial_{z_{1}} \partial_{\bar{z}_{1}}
G(z_{1},z_{2})\partial_{{z}_{2}}
G(z_{1},z_{2}) \partial_{\bar{z}_{2}}G(z_{1}, z_{2}) \\

a_6=\int \left[ dm \right]_2^{1} e^{\alpha ^{\prime} p^{(1)} \cdot p^{(2)}
G(z_{1},z_{2}) }|z_{1}|^{2} |z_{2}|^{2}\partial_{z_{1}} \partial_{\bar{z}_{2}}
G(z_{1},z_{2}) \partial_{\bar{z}_{1}}
G(z_{1},z_{2}) \partial_{z_{2}}  G(z_{1},z_{2}) \\

a_7=\int \left[ dm \right]_2^{1} e^{\alpha ^{\prime} p^{(1)} \cdot p^{(2)}
G(z_{1},z_{2}) }|z_{1}|^{2} |z_{2}|^{2}\partial_{\bar{z}_{1}} \partial_{{z}_{2}}
G(z_{1},z_{2}) 
\partial_{z_{1}} 
G(z_{1},z_{2}) \partial_{\bar{z}_{2}} G(z_{1},z_{2}) \\

a_8=\int \left[ dm \right]_2^{1} e^{\alpha ^{\prime} p^{(1)} \cdot p^{(2)}
G(z_{1},z_{2}) }|z_{1}|^{2} |z_{2}|^{2}\partial_{z_{2}} \partial_{\bar{z}_{2}}
G(z_{1},z_{2}) \partial_{z_{1}}
G(z_{1},z_{2})  \partial_{\bar{z}_{1}} G(z_{1},z_{2}) \\

a_9=\int \left[ dm \right]_2^{1} e^{\alpha ^{\prime} p^{(1)} \cdot p^{(2)}
G(z_{1},z_{2}) }|z_{1}|^{2} |z_{2}|^{2} \partial_{\bar{z}_{1}} 
\partial_{\bar{z}_{2}}
G(z_{1},z_{2}) \partial_{{z}_{1}}
G(z_{1},z_{2}) \partial_{z_{2}}  G(z_{1},z_{2}) \\

a_{10}=\int \left[ dm \right]_2^{1} e^{\alpha ^{\prime} p^{(1)} \cdot p^{(2)}
G(z_{1},z_{2}) }|z_{1}|^{2} |z_{2}|^{2} \partial_{{z}_{1}}G(z_{1},z_{2})
\partial_{\bar{z}_{1}} G(z_{1},z_{2}) \partial_{{z}_{2}}
G(z_{1},z_{2}) \partial_{\bar{z}_{2}}  G(z_{1},z_{2})

\end{array}
$$

While these coefficients can be computed in the limit
$ \alpha' \rightarrow 0 $ (and
we will give the results in a forthcoming
paper), some of their properties can be proved on general grounds. 
By integrating by parts all the terms containing double derivatives of the
Green function, it can be shown that only three $a_{i}$'s are independent.
We would like here to stress that at the one-loop order it is possible,
due to the properties of the Green function on the torus, to disregard
surface terms coming from the integration by parts. Hence this case
results to be even easier than the tree level. 

Indeed, at one-loop order the expression for $T_{\mu \nu \rho \sigma}$ drastically simplifies
reducing to:
\beqa
T_{\mu \nu \rho \sigma} &  =  & \bigg\{ - \frac{2}{p^{2}} (a_{3} + a_{2} )
\left[ \eta_{\mu \rho} p_{\nu} p_{\sigma} + \eta_{\nu \rho} p_{\mu} p_{\sigma}
+ \eta_{\mu \sigma} p_{\rho} p_{\nu} + \eta_{\nu \sigma} p_{\rho} p_{\mu}
\right]  \nonumber \\
& {} & - \frac{4}{p^{2}} a_{1} \left[ \eta_{\mu \nu} p_{\rho} p_{\sigma} +
\eta_{\rho \sigma} p_{\mu} p_{\nu} \right] + \frac{4}{p^{4}} (a_{1}+a_{2}
+a_{3}) p_{\mu} p_{\nu} p_{\rho} p_{\sigma} \nonumber \\
& {} & +  2 (a_{3}+a_{2}) \left[ \eta_{\mu \rho} \eta_{\nu \sigma} +
\eta_{\nu \rho} \eta_{\mu \sigma} \right] + 4 a_{1} \eta_{\mu \nu}
\eta_{\rho \sigma} \bigg\} \nonumber \\
& {} & + \bigg\{ - \frac{2}{p^{2}} (a_{3} - a_{2} ) \left[ \eta_{\mu \rho}
p_{\nu} p_{\sigma} - \eta_{\nu \rho} p_{\mu}p_{\sigma} + \eta_{\nu \sigma}
p_{\mu} p_{\rho} - \eta_{\mu \sigma} p_{\rho} p_{\nu} \right] \nonumber \\
& {} & + 2 (a_{3} - a_{2} )  \left[ \eta_{\mu \rho} \eta_{\nu \sigma}
- \eta_{\mu \sigma} \eta_{\nu \rho} \right] \bigg\} \ \equiv \ S_{\mu \nu
\rho \sigma} + A_{\mu \nu \rho \sigma} 
\eeqa
where we used the momentum conservation for setting $p^{(1)}=-p^{(2)} \equiv p$
and made explicit the symmetry properties on the indices $(\mu \nu)$ which
refer to the polarization tensor of the particle (1)  and 
$(\rho \sigma)$ which refer to the one of the particle (2). Of course
the amplitude is symmetric under the exchange of the states $(1)$ and
$(2)$. 

We notice that the symmetric part $S_{\mu \nu \rho \sigma}$ depends only on
the two quantities $a_{1}$ and $a_{2}+a_{3}$, while the antisymmetric one 
$A_{\mu \nu \rho \sigma}$ has an overall factor $a_{3}-a_{2}$. 

The one loop amplitude, {\em as an off-shell string amplitude}, has the very 
remarkable property of being transverse:
\[
p^{\mu} T_{\mu \nu \rho \sigma} = 0  .
\]
  
In this respect the relations among the quantities $a_{i}$'s play the role
of Ward identities for the two-point function, which in field theory 
follow from the gauge invariance (i.e. invariance under general coordinates
transformations). However in quantum gravity 
transversality does not always hold;
it does, for instance, when background-field techniques are used~\cite{D}.

In the final part of this work we will check that
our prescription (\ref{GAUGE}) also works in the open string case~\cite{DMLRM}.

The starting point is given by the $N$-string $g$-loop Vertex
for the open string~\cite{DPFHLS}:
\[
V_{N;g}= {\cal C}_{g}^{open} \prod_{i=1}^{N} \bigg[ {}_i\!<\!
p_{i};O_{a}|\bigg] \,\,\delta  \left( \sum_{i=1}^{N} p_{i} \right)
 \int [{\mbox d}m]^{N}_{g}
\,{\hat{V}}_{N;0}
\]
\[
\times \exp\left\{-\frac{1}{2}\sum_{i,j=1}^{N}
\oint_0 dz \oint_0 dy \partial X^{(i)}(z)
\log
\frac{ E\left(V_i(z),V_j(y)\right)}
{V_{i}(z)-V_{j}(y)}
\partial X^{(j)}(y)
\right\}
\]
\begin{equation}
\times \!\int\! \prod_{\mu=1}^{g} [ d^{D} k_{\mu} ] \exp\!\left\{\!
-\alpha'\!  \sum_{\mu,\nu=1}^{g} k_{\mu} (2 \pi {\mbox Im} \tau )_{\mu\nu} k_{\nu}
+
i \sqrt{2 \alpha'} \sum_{i=1}^{N} \sum_{\mu=1}^{g}k_{\mu}\!
\left[\oint_0 dz \partial X^{(i)}(z)\!
\left(\!
\int_{z_{0}}^{V_{i}(z)}\! \omega^{\mu}\! \right) \right]\!
\right\}
\label{VNGO}
\eeq

\vspace{0.5cm}
with $\alpha_{0}^{\mu}=\sqrt{2 \alpha'} p^{\mu}$ and $\hat{V}_{N;0}$
defined in~(\ref{VN0}). From~(\ref{VNGO}) one can have the corresponding
expression for $N$ external ``photons''. It is given by:
\[
{\cal V}_{N;g}\, =\, {\cal C}_{g}^{open}\! < \Omega| \int [ {\mbox d} m]^{g}_{N}
\exp \left\{\frac{1}{2} \sum_{i=1}^{N} \sqrt{2 \alpha'} p^{(i)} \cdot
\left.\! \left[ \sqrt{2 \alpha'} p^{(i)} + {\alpha}_{1}^{(i)} 
\partial_{z} \right] \log V'_{i}(z)
\right|_{z=0} \right\}
\]
\begin{equation}
\times \exp \frac{1}{2} \bigg\{ \sum_{\stackrel{i,j=1}{i \neq j}}^{N} 
\left[ \sqrt{2 \alpha'} p^{(i)} + \alpha_{1}^{(i)} V'_{i}(0)
\partial_{z_{i}}\right]
\cdot \left[ \sqrt{2 \alpha'} p^{(j)} + \alpha_{1}^{(j)}
V'_{j}(0) \partial_{z_{j}} \right] {\cal G}(z_{i},z_{j} ) \bigg\}
\label{VMP}
\end{equation}
where
\beq
 {\cal G}(z_{i},z_{j}) \equiv  \log E(z_{i},z_{j})
- \frac{1}{2}  \left(
\int_{z_{i}}^{z_{j}} \omega^{\mu} \right) ( 2 \pi {\mbox Im} \tau
 )_{\mu \nu}^{-1} \left( \int_{z_{i}}^{z_{j}} \omega^{\nu} 
\right)           \label{GFO}
\eeq
is the $g$-loop Green function that, also in this case, can be rescaled
as:
\beq
 G(z_{i},z_{j}) \equiv  \log \frac{E(z_{i},z_{j})}{\sqrt{V'_{i}(0)
 V'_{j}(0)}}
- \frac{1}{2}  \left(
\int_{z_{i}}^{z_{j}} \omega^{\mu} \right) ( 2 \pi {\mbox Im} \tau
 )_{\mu \nu}^{-1} \left( \int_{z_{i}}^{z_{j}} \omega^{\nu} 
\right) \label{GV}.            
\eeq

Once again we restrict ourself to the $g=1$ loop case. The choice (\ref{GAUGE}) in (\ref{GV})
reproduces the right translational invariant Green function and allows us 
to rewrite the Vertex
${\cal V}_{N;1}$ only in terms of it:
\[
{\cal V}_{N;1}\, =\, {\cal C}_{1}^{open}\! < \Omega| \int [ {\mbox d} m]^{1}_{N}
\]
\begin{equation}
\times \exp \bigg\{ \sum_{\stackrel{i,j=1}{i \neq j}}^{N} 
\left[ \sqrt{2 \alpha'} p^{(i)} + \alpha_{1}^{(i)} z_{i}
\partial_{z_{i}} \right]
\left[ \sqrt{2 \alpha'} p^{(j)} + \alpha_{1}^{(j)}
z_{j} \partial_{z_{j}} \right]  G(z_{i},z_{j} ) \bigg\} .
\label{VMP1}
\end{equation}
By saturating ${\cal V}_{N;1}$ on two photon states, defined by:
\[
| \epsilon; p > = {\cal N}^{ph.}_{0} \epsilon_{\mu} \alpha^{\mu}_{-1} |p>
\]
we get the corresponding amplitude:
\begin{equation}
A_{2;1} = ({\cal N}^{ph.}_{0})^{2} {\cal C}^{open}_{1} 
\epsilon^{(1)\mu} \epsilon^{(2)\nu}
T_{\mu \nu}
\end{equation}with
\[
T_{\mu \nu} = 2 \left( a_{1} \eta_{\mu \nu} + 4 \alpha^{'} a_{2} p^{(1)}_{\nu}
p^{(2)}_{\mu} \right)
\]
where the coefficients $a_{i}$'s are given by:

$$
\begin{array}{ll}
a_1=\int \left[ dm \right]_{2}^{1} e^{4{\alpha}^{\prime} p^{(1)} \cdot p^{(2)}
G(z_{1},z_{2})}z_{1} z_{2} \partial_{z_{1}} \partial_{z_{2}} 
G(z_{1},z_{2}) \\

a_2=\int \left[ dm \right]_2^{1} e^{ 4 \alpha ^{\prime} p^{(1)} \cdot p^{(2)}
G(z_{1},z_{2}) } z_{1} z_{2} \partial_{z_{1}} G_(z_{1},z_{2}) 
\partial_{{z}_{2}} G(z_{1},z_{2}) .

\end{array}
$$
By integrating by parts the term containing the double derivative of 
$G(z_{i},z_{j})$,  we get a relation between $a_{1}$ and $a_{2}$:
\[
a_{1}= - 4 \alpha' p^{(1)} \cdot p^{(2)} a_{2}
\]
so that we derive for $T_{\mu \nu}$ the following expression:
\begin{equation}
T_{\mu \nu} = 4 \alpha' a_{2} \left[ p^{2} \eta_{\mu \nu} - p_{\nu} p_{\mu}
\right]
\end{equation}

Also this amplitude, even being an off-shell string amplitude, 
is transverse. Once
$a_2$ is evaluated in the $\alpha' \rightarrow 0$ limit, the results
of gauge field theory are properly reproduced \cite{DMLRM}.

In conclusion, we have given here a systematic way of obtaining off-shell
one-loop amplitudes with $N$ external massless states; 
the same procedure can be now applied to the Klein bottle \cite{CMPP2}. 

We defer to a forthcoming work a full analysis of the low energy limit
of off-shell both of open and closed string amplitudes.

\vspace{1cm}

ACKNOWLEDGMENTS. We thank P. Di Vecchia for many helpful discussions and for 
useful comments on a preliminary version of the paper.

\end{document}